\documentclass[10pt, conference, letterpaper]{IEEEtran}
\usepackage[utf8]{inputenc}
\usepackage[T1]{fontenc}
\usepackage{amsmath, amssymb, amsfonts}
\usepackage{graphicx}
\usepackage{booktabs}
\usepackage{hyperref}

\title{Premature Dimensional Collapse and Tensor-based Execution Paths for High-Dimensional Relational Operations in Cost-Based Database Systems}

\author{\IEEEauthorblockN{Il-Sun Chang}
\IEEEauthorblockA{\textit{Independent Researcher} \\
Daegu, Republic of Korea \\
kyou0072@gmail.com}
}

\begin{document}

\maketitle

\begin{abstract}
This study structurally reinterprets the phenomenon of "premature dimensional collapse" that occurs during high-dimensional relational operations in cost-based database systems and empirically analyzes its impact on performance characteristics at the point of data representation decision in the execution layer. Existing relational execution paths flatten data at an early stage of operation and rely on hash-based structures to perform joins. While this approach is efficient in environments with sufficient memory resources, under resource constraints, it triggers repetitive partitioning and spilling of hash tables, leading to a non-linear increase in execution costs and a sharp expansion of tail latency. Experimental results show that under the condition of input scale N=1,000,000 and work\_mem=1MB, the relational execution path generated approximately 200MB (25,662 blocks) of temporary disk I/O, with P99 tail latency exceeding 2 seconds. In contrast, the tensor-based execution path maintained a stable P99 latency of approximately 0.56 seconds without physical spilling under the same conditions. This difference becomes more evident when considering not only the average execution time (P50) but also the tail latency (P99) and physical I/O metrics (Temp\_MB) together. This study does not merely propose operational acceleration but demonstrates that the timing of data representation decisions in the execution layer can affect the structural stability of performance. While relational execution loses predictability under memory pressure, tensor-based execution maintains more deterministic scaling characteristics. This suggests that designing execution paths that delay or bypass premature dimensional collapse in resource-constrained environments can play a crucial role in stabilizing tail latency and reducing structural recovery costs.
\end{abstract}

\section{Introduction}
Cost-based database systems determine the execution method of major operations, including joins and sorts, in advance during the query optimization stage and follow a structure that efficiently performs the plan during the execution stage. This separation has served as a core design principle to make the processing of complex queries manageable, and the selection of join algorithms or sorting methods is traditionally fixed during optimization based on cost estimation results. However, in modern workload environments dealing with large-scale data and complex join structures, cases where such early decisions lead to severe performance bottlenecks during the execution stage are being observed more frequently. In particular, large-scale hash joins and sort operations are the operations that involve the most memory access and intermediate result generation in relational execution engines, and a significant portion of the execution cost arises from data movement and representation transformation processes rather than the computation of the operation itself. These costs expand non-linearly as data scale increases and are difficult to explain sufficiently by simple cost estimation errors or the inefficiency of specific algorithms alone. This is because existing cost-based optimization has focused mainly on comparing costs at the operation unit, while failing to sufficiently address the issue of when and in what form data representation and execution paths are fixed.

In relational databases, join and sort operations can essentially be interpreted as high-dimensional relational operations where multiple attributes and join conditions act simultaneously. Nevertheless, existing execution engines have adopted a method of reducing these high-dimensional structures to linearized intermediate representations at an early stage of execution. Such premature dimensional reduction simplifies execution at the cost of losing the join information inherent in the relational structure, triggering unnecessary intermediate result expansion and repetitive memory movement in subsequent execution stages. Consequently, inefficient execution paths are forced without sufficiently reflecting the actual data distribution or join characteristics observed during the execution stage.

This study defines the fundamental cause of these performance bottlenecks not as the computational complexity of the operation itself, but as the "premature fixation of decision points across layers and the resulting dimensional collapse" present throughout cost-based database systems. Since data representation and execution paths are pre-confirmed during the query optimization stage, the room for reflecting additional information obtained at the time of execution into the execution strategy is limited, appearing as structural inefficiencies in large-scale join and sort operations. These problems are difficult to solve fundamentally through fine-grained optimization inside the execution engine alone. From this perspective, this paper proposes a tensor-based execution path that complements the existing CPU-based linear execution path for large-scale hash joins and sort operations. The proposed approach aims to delay the point where linearization and materialization become inevitable until the late stages of execution by performing operations while maintaining the multidimensional structure of data as long as possible. Tensor operations naturally support such dimension-preserving calculations and are advantageous for reducing repetitive memory movement and intermediate result generation, especially in large-scale datasets. Furthermore, this study does not apply tensor-based execution collectively to all operations. Instead, it introduces a lightweight decision policy that selects the operation path at the time of execution based on data scale and structural coupling, selectively applying tensor-based execution only when it is structurally advantageous. Through this, it is shown that the problem of dimensional collapse caused by early decisions can be mitigated in the execution layer while maintaining compatibility with existing cost-based optimization frameworks.

Recent studies have explored GPU-accelerated relational processing and column-oriented execution models to mitigate memory overhead and improve throughput \cite{bress2014, he2008, abadi2007}. While these systems primarily target throughput improvement or operator-level acceleration, they do not explicitly address execution-time representation stability under memory-regime transitions. In contrast, this work addresses the structural cost implications of premature representation decisions in cost-based execution itself, and analyzes predictability degradation under memory-regime shifts. In particular, we argue that performance instability under memory pressure is fundamentally a predictability problem rather than a pure throughput limitation.

The main contributions of this paper are as follows: First, performance bottlenecks occurring in large-scale join and sort operations are reinterpreted from the perspective of high-dimensional relational operations, and the impact of premature dimensional collapse and decision point issues on execution costs is analyzed. Second, a tensor-based hash join and sort execution path is presented to mitigate these problems, and the structural differences from existing linear execution methods are explained. Third, through experiments in CPU and GPU environments, it is demonstrated that the proposed approach effectively reduces memory access costs and tail latency under specific conditions.

\section{Background}
\subsection{Cost-based Execution Model and Decision Points}
Traditional cost-based database systems explore possible execution plan candidates during the query optimization stage and select one execution plan through cost estimation. In this process, major execution decisions such as join order, join algorithm, and sorting method are confirmed in advance during the optimization stage, and the primary role of the execution stage becomes performing the selected plan as efficiently as possible. While this design has contributed to securing simplicity and predictability in the execution stage, it creates a clear disconnect in the decision point between the execution stage and the optimization stage. Decisions made during optimization are maintained regardless of dynamic information such as actual data distribution, intermediate result size, and memory pressure observed at execution time, and it is difficult to reverse the results of fixed decisions in the execution stage. Especially in large-scale data environments, the premature fixation of these decision points itself frequently has a greater impact on execution efficiency than errors in cost estimation.

\subsection{High-Dimensional Relational Operations and Linear Execution Models}
Join and sort operations in relational databases are operations that are difficult to reduce to simple record-unit comparisons or sequential processing. Join operations involve join conditions for multiple attributes acting simultaneously, and sort operations also include the process of mapping order relationships in multidimensional attribute spaces to linear orders. From this perspective, joins and sorts can be interpreted as operations that essentially handle high-dimensional relational structures. Nevertheless, existing relational execution engines process these high-dimensional structures by converting them into 1D or low-dimensional linear representations at an early stage of execution. For example, a hash join converts the input relation into a linear memory structure called a hash table, and a sort operation rearranges the multidimensional key space into a linear order based on a single comparison criterion. While this linearization simplifies execution and enables the application of general-purpose algorithms, it comes at the cost of prematurely erasing the structural join information of the data.

\subsection{Premature Dimensional Collapse and Execution Costs}
The phenomenon occurring as high-dimensional relational structures are linearized in the early stages of execution is defined in this paper as "premature dimensional collapse". Premature dimensional collapse does not simply mean a transformation of data representation but is accompanied by a structural loss where the join possibilities and constraints inherent in the relational structure can no longer be utilized during the execution process. This dimensional collapse directly affects execution costs. Linearized intermediate representations often generate intermediate results larger than actually necessary, leading to increased memory usage and repetitive data movement. Particularly in hash joins and sort operations, the size of intermediate results and memory access patterns have a decisive impact on execution time and tail latency, and these costs become a more dominant factor than the computational complexity of the operation itself. The important point is that this cost increase does not stem from errors in selecting specific algorithms or inefficiencies in implementation, but from a structural design where the decision point is moved too far forward. Even if additional information is secured during the execution stage, already linearized representations and fixed execution paths limit the room to utilize such information. This phenomenon appears as a structural problem where the linear execution path fails to maintain ideal linear scaling (O(N)) and deviates sharply as the input scale increases.

\begin{figure}[h]
    \centering
    \includegraphics[width=0.9\linewidth]{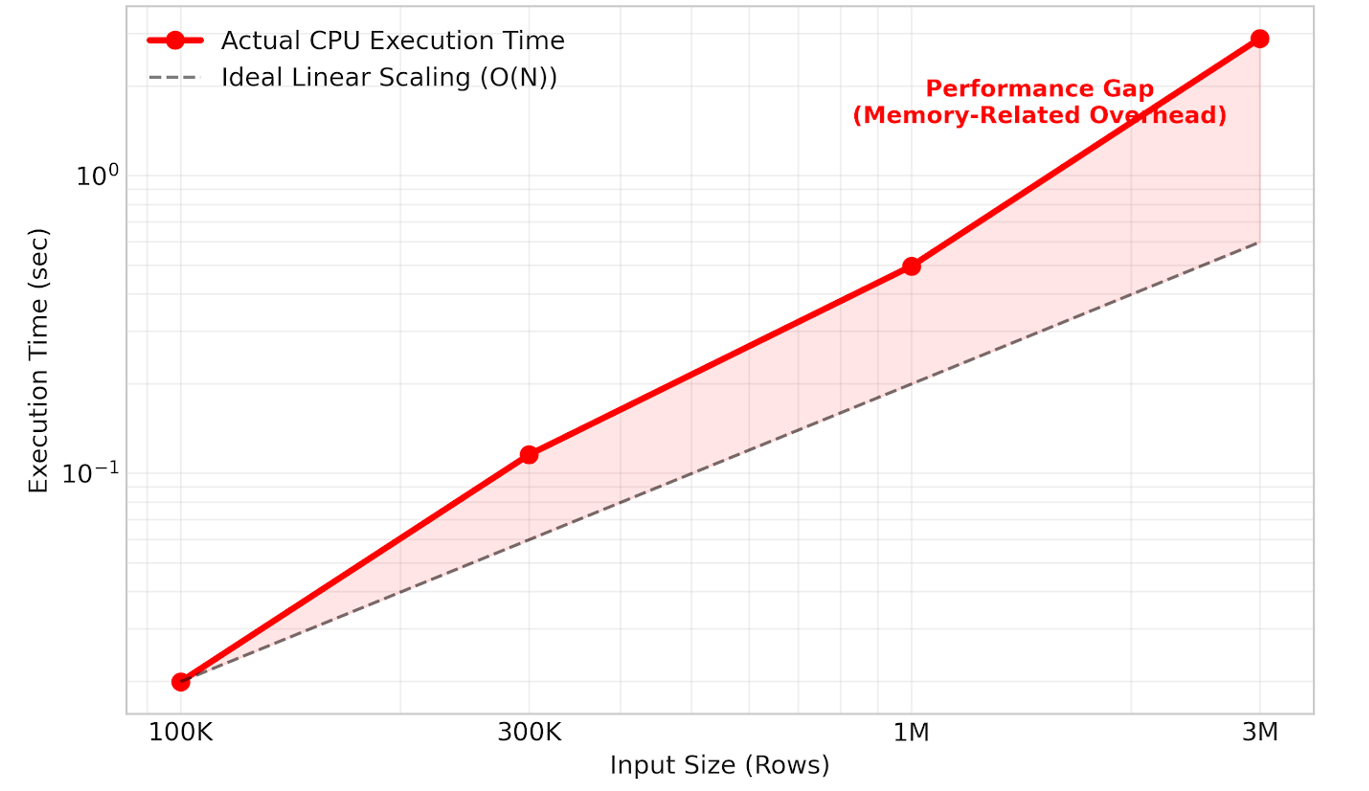}
    \caption{Scalability collapse of CPU Hash Join}
\end{figure}

\subsection{Possibility of Alternative Approaches in the Execution Layer}
Existing cost-based optimization has mainly developed in the direction of improving the accuracy of plan selection during the optimization stage. On the other hand, approaches that reconsider the data representation and operation structure itself in the execution layer have been handled relatively limitedly. However, in cases where premature dimensional collapse acts as the primary cause of execution costs, providing an alternative execution path that considers dimension preservation in the execution layer can be structurally more effective. In this context, tensor operations, which can more directly represent and operate on high-dimensional relational structures, can be considered a meaningful alternative in the execution layer. Tensor operations can naturally represent multidimensional data structures and offer the possibility of mitigating intermediate result expansion and unnecessary memory movement by delaying the point where linearization becomes inevitable. Based on this perspective, this paper introduces a tensor-based execution path as a complementary option in the execution layer for large-scale hash joins and sort operations.

\begin{figure}[h]
    \centering
    \includegraphics[width=0.9\linewidth]{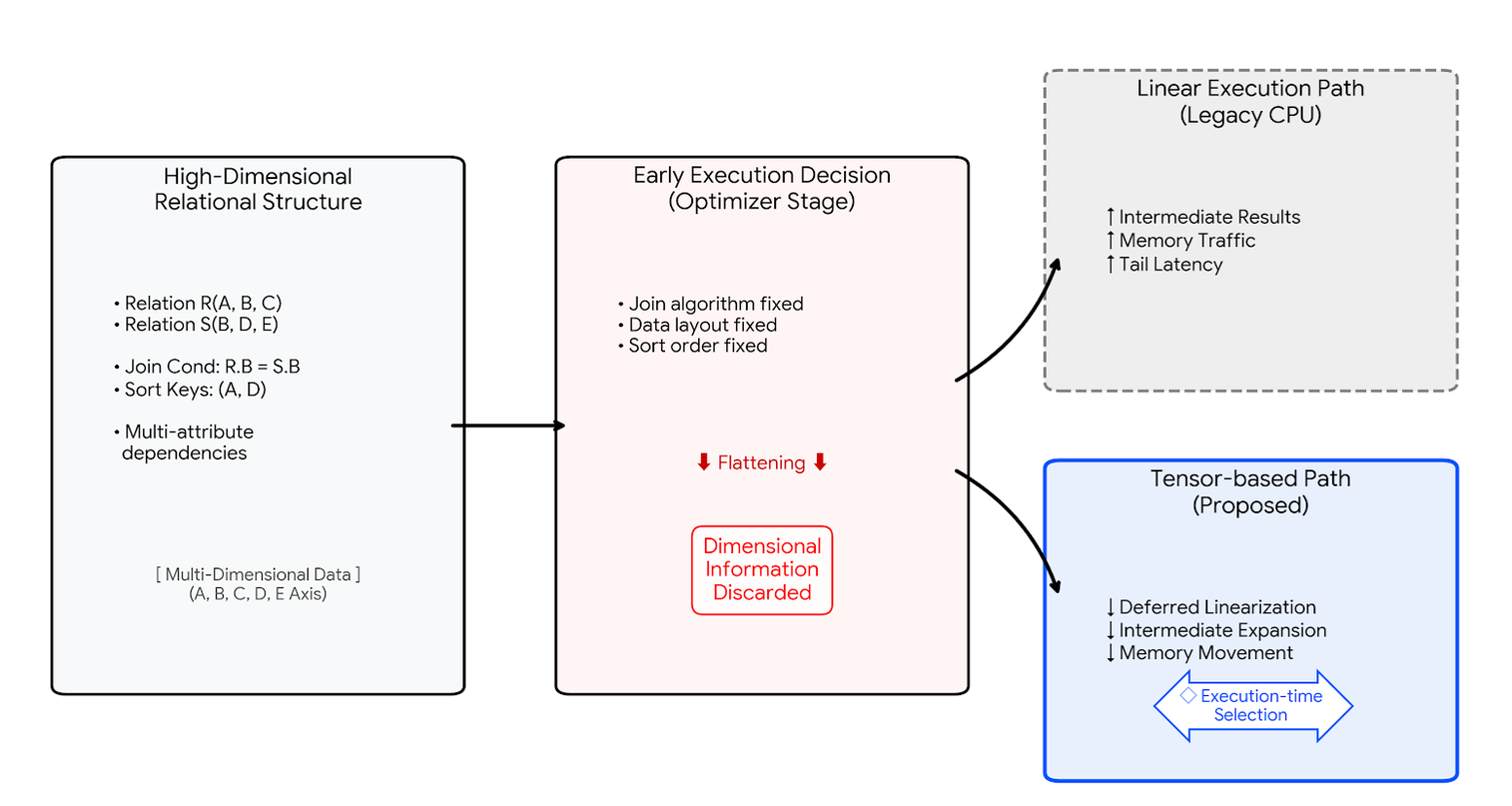}
    \caption{Premature dimensional collapse caused by early execution decisions}
\end{figure}

\section{Tensor-based Execution Model – Overview}
\subsection{Overview}
As seen in Figure 2, the performance bottleneck of large-scale join and sort operations occurring in cost-based database systems originates primarily from premature dimensional collapse, where data representation and execution paths are linearized in the early stages of execution, rather than the computational complexity of the operations themselves. This problem is difficult to mitigate sufficiently only by improving the accuracy of cost estimation at the optimization stage, and an alternative execution path that can handle data structures more flexibly in the execution layer is required. The tensor-based execution model proposed in this paper aims to provide a "complementary execution path" for large-scale hash joins and sort operations rather than replacing existing linear execution paths.

The core idea lies in maintaining relational operations in high-dimensional structures as much as possible until the point of execution and delaying the time when linearization and materialization are actually needed. Through this, unnecessary expansion of intermediate results and repetitive memory movement can be mitigated, and data characteristics observed at the execution stage can be more directly reflected. In the tensor-based execution model, relational data is mapped to multidimensional tensor representations, and join and sort operations are performed on these tensor structures. At this time, the tensor representation is used not simply as a means for GPU acceleration, but as a representation means to naturally preserve the relational structure where multiple attributes and join conditions act simultaneously. In other words, this approach is closer to a structural alternative that reconsiders the choice of data representation in the execution layer rather than an acceleration technique dependent on specific hardware.

Crucially, this model does not force tensor-based execution for all operations. Tensor-based execution is selectively applied only when the data scale is large enough and the coupling of the relational structure is high, such that the cost due to dimensional collapse in the linear execution path becomes prominent. To this end, a simple condition check is performed at execution time to select between the existing CPU-based linear execution path and the tensor-based execution path. This execution-time selection is designed to supplement inefficiencies caused by early decisions in the execution layer while minimizing conflict with existing cost-based optimization frameworks. This tensor-based execution model provides a practical approach to mitigate structural bottlenecks occurring in large-scale hash joins and sort operations without changing the overall structure of existing relational databases.

\subsection{Tensor Representation of Relational Data}
The starting point of the tensor-based execution model lies in the way relational data is represented in tensor form. In existing relational execution engines, input tables are treated as sets of tuples or linear memory layouts, and join and sort operations are performed on the premise of these linear representations. In contrast, this study interprets relational data as a multidimensional structure and introduces tensor form as a representation that can maintain this in the execution layer. A relational table consists of multiple attributes, and each attribute can be considered an independent dimension. For example, a relation R(A, B, C) can be interpreted as a multidimensional data structure with axes corresponding to attributes (A, B, C). In this case, each tuple is represented as a coordinate on the corresponding axes, and the entire table can be seen as a sparse multidimensional space formed by these sets of coordinates. The tensor representation used in this paper aims to maintain the multi-attribute structure of relational data until the time of execution based on this perspective.

In the case of join operations, join conditions for common attributes are interpreted as alignment or axis matching problems between tensor axes. For example, a join between R(A, B, C) and S(B, D, E) can be expressed as an operation that aligns and joins two tensors based on the axis corresponding to attribute B. In this representation, the join condition is not reduced to a linearized comparison operation but is explicitly maintained within the multidimensional structure. As a result, structural information related to the join condition is not lost in the intermediate representation stage. Sort operations can also be interpreted in a similar way. When using multi-attribute sort keys, existing execution engines process them by converting comparisons for multiple attributes into a single linear order. In contrast, in tensor-based representation, relocation can be performed on the attribute axes corresponding to the sort keys while maintaining them explicitly. This offers the possibility of reducing unnecessary intermediate data rearrangement and memory movement during the sorting process.

The important point is that this tensor representation does not mean converting all data into dense high-dimensional arrays. In actual implementation, sparse tensor representations or block-unit partial tensor representations can be utilized depending on the distribution and density of data, which is compatible with the physical storage structure of existing relational data. The focus of this study is not on the choice of a specific tensor representation method, but on delaying premature dimensional collapse by introducing a representation that can maintain the high-dimensional structure of relational data in the execution layer.

\subsection{Execution-time Path Selection}
The core of the tensor-based execution model is not the tensor operation itself, but "delaying the decision point so that the operation path can be selected at the time of execution". In existing cost-based database systems, major execution decisions such as join algorithms, data layouts, and sorting methods are fixed in advance during the optimization stage, and it is difficult to change these decisions during the execution stage. This study maintains this structure but provides alternative execution paths selectable in the execution layer only for large-scale join and sort operations. Execution-time selection does not fundamentally change the execution plan generated during the optimization stage. Instead, for operations with a large variance in execution costs such as hash joins and sorts, it selects either a linear execution path or a tensor-based execution path using limited information observed during the execution stage.

The judgment criteria used for execution path selection are intentionally designed to be simple. This paper utilizes indicators that are relatively easy to observe at the time of execution, such as the scale of input data, the cardinality of join attributes, and the expected intermediate result size. These indicators are not intended to replace accurate cost estimation but are used as signals to identify cases where the possibility of cost increase due to premature dimensional collapse in the linear execution path is high. Crucially, execution-time selection does not change the semantic result of the operation or the consistency of the query.

\section{Tensor-based Hash Join and Sort}
\subsection{Tensor-based Hash Join}
Existing hash join algorithms are performed by converting input relations into linear memory structures and then constructing and searching hash tables based on join keys. In this process, structural relationships between attributes other than the join key are erased in the early stages of execution, and intermediate results are generated in a flattened form based on the join key. Especially in large-scale data environments, the size of hash tables and memory access during the search process act as major factors in execution costs. In a tensor-based hash join, input relations are mapped to multidimensional tensor representations. At this time, attributes corresponding to join keys are aligned on a common axis, and other attributes are maintained as independent axes. The join operation is defined in the form of performing a combination based on the common axis on this tensor representation.

\subsection{Tensor-based Sort}
The sort operation is the process of rearranging relational data into a linear order based on multi-attribute keys, and like joins, it reduces high-dimensional structures into a single linear order at an early stage of execution. Existing execution engines use comparison-based sorting algorithms or external sorting techniques, during which repetitive data movement and memory access occur. Particularly, sorting large-scale data has a significant impact not only on execution time but also on tail latency. In a tensor-based sort, relocation is performed on the tensor axes while maintaining multi-attribute sort keys explicitly. In other words, combinations of attributes corresponding to sort keys are not immediately reduced to linear comparison operations but are sorted step-by-step within the multidimensional structure. This approach offers the possibility of reducing unnecessary global rearrangement and maintaining memory access patterns more regularly during the sorting process.

\section{Experimental Evaluation}
\subsection{Experimental Setup}
All experiments were conducted based on a prototype-level implementation. The linear execution path uses standard CPU-based hash join and sort algorithms, while the tensor-based execution path was implemented to perform join and sort operations on multidimensional tensor representations. Both execution paths use the same input data and logical queries, and the consistency of the results was always confirmed. Experiments were conducted in CPU and GPU environments respectively.

\subsection{Hash Join Performance}
In the interval where the data scale is small, the existing CPU-based hash join showed a lower execution time than the tensor-based execution. However, in intervals where the input data scale increases and the expansion of intermediate results becomes prominent due to join key cardinality, intervals where the tensor-based hash join shows a lower execution time than the linear execution path were observed. Figure 3 shows how the peak memory usage of the internal hash table increases in CPU hash joins as the input size increases.

\begin{figure}[h]
    \centering
    \includegraphics[width=0.9\linewidth]{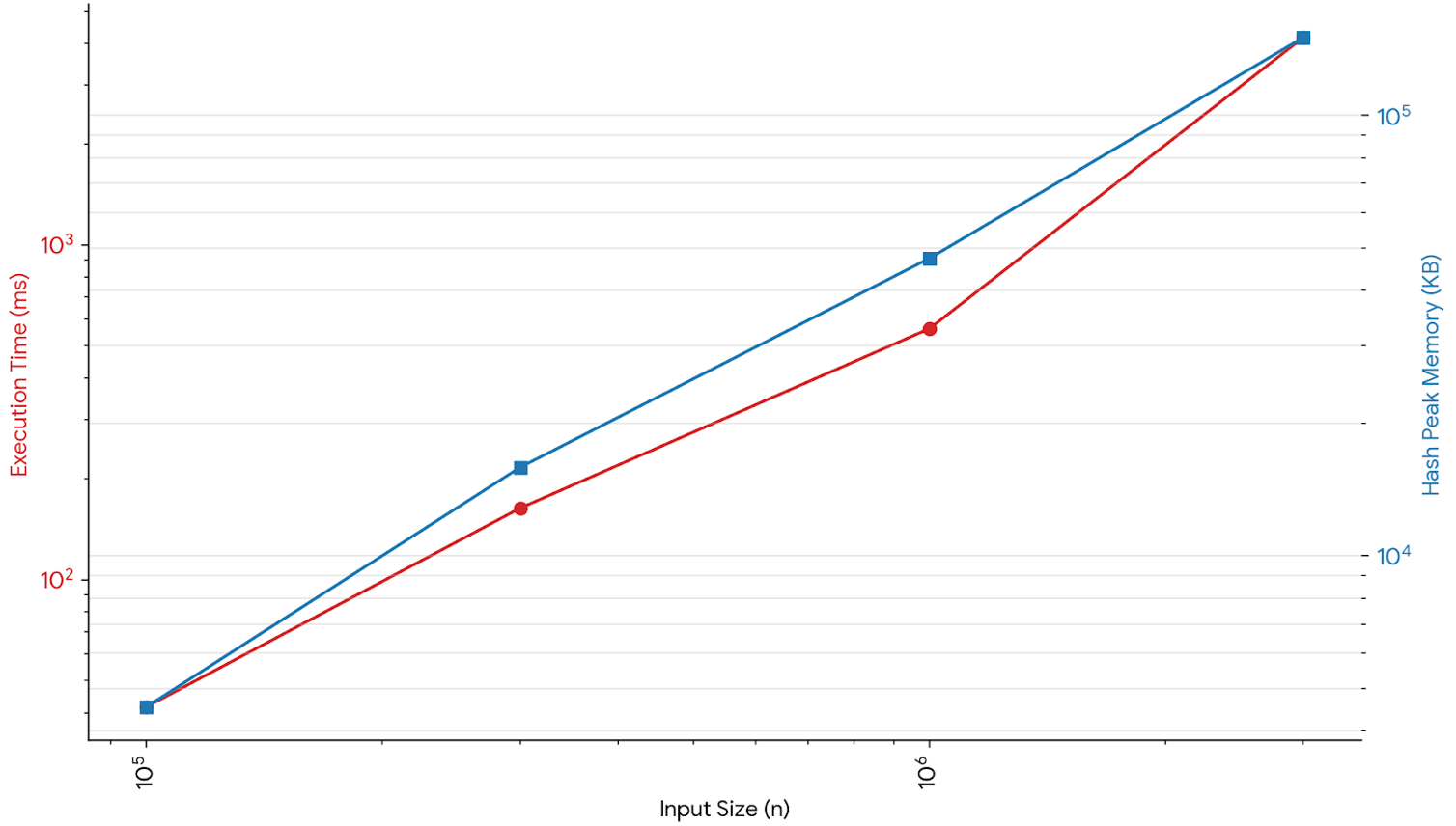}
    \caption{Growth of Intermediate Hash Table Due to Premature Linearization}
\end{figure}

In terms of tail latency, a similar trend was observed. For linear hash joins, the variance of the latency distribution expanded significantly as the input scale increased, whereas the tensor-based execution path maintained a stable distribution. Figure 4 shows the latency distribution of CPU hash joins. P99 and maximum latency increase significantly compared to the median as the input scale increases.

\begin{figure}[h]
    \centering
    \includegraphics[width=0.9\linewidth]{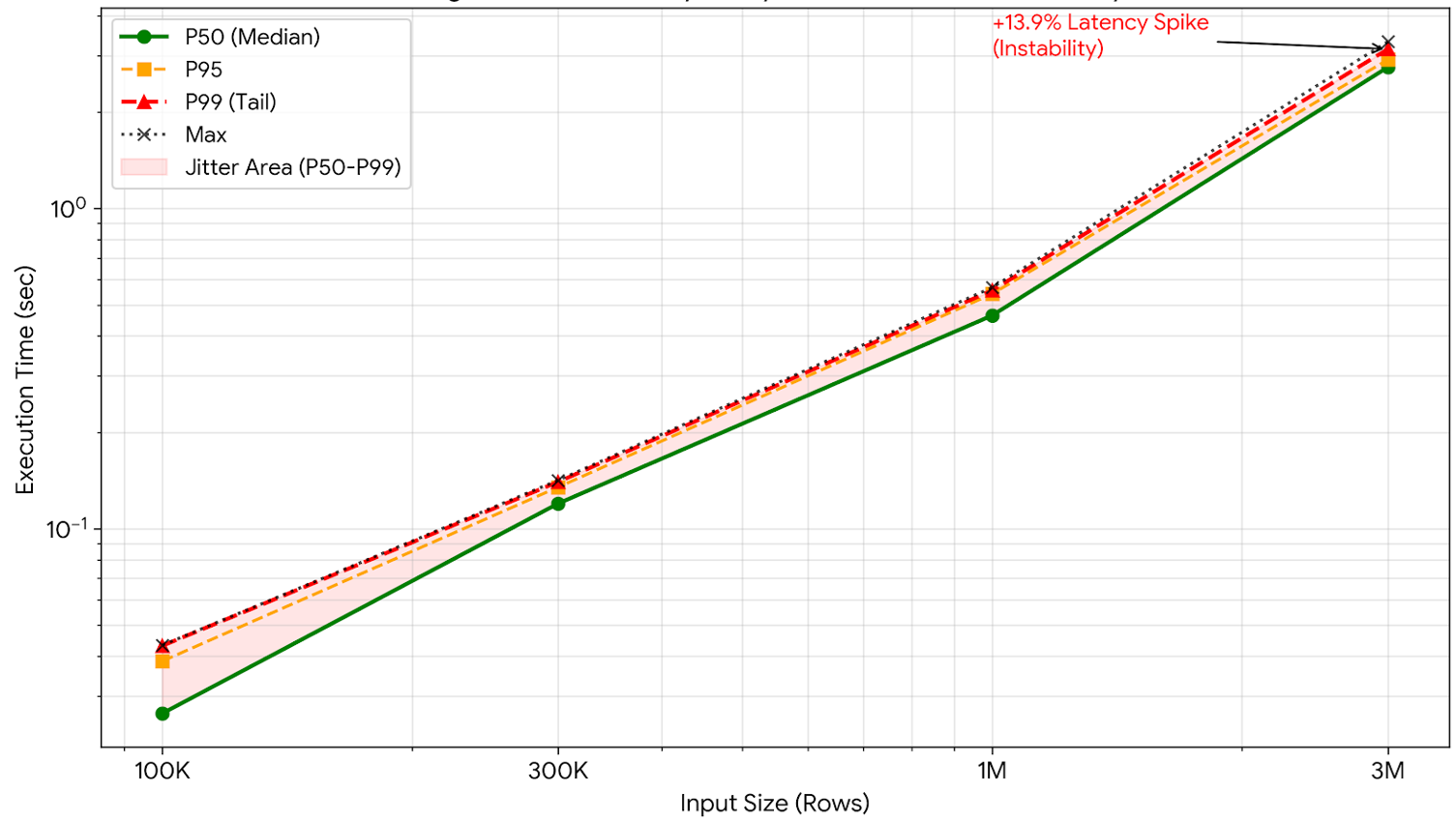}
    \caption{Tail Latency Analysis: CPU Execution Stability}
\end{figure}

\subsection{Sort Performance}
Evaluation of the sort operation was conducted primarily on scenarios using multi-attribute sort keys. In single-key sorts or small datasets, the existing linear sort path showed lower execution time. Conversely, as the data scale grew and the dimension of sort keys increased, repetitive data rearrangement and memory movement occurring in the linear sort path acted as major factors in execution costs. In these intervals, the tensor-based sort tended to mitigate execution time and tail latency by reducing intermediate data movement. Figure 5 shows the execution time difference between single-key and multi-attribute sorts.

\begin{figure}[h]
    \centering
    \includegraphics[width=0.9\linewidth]{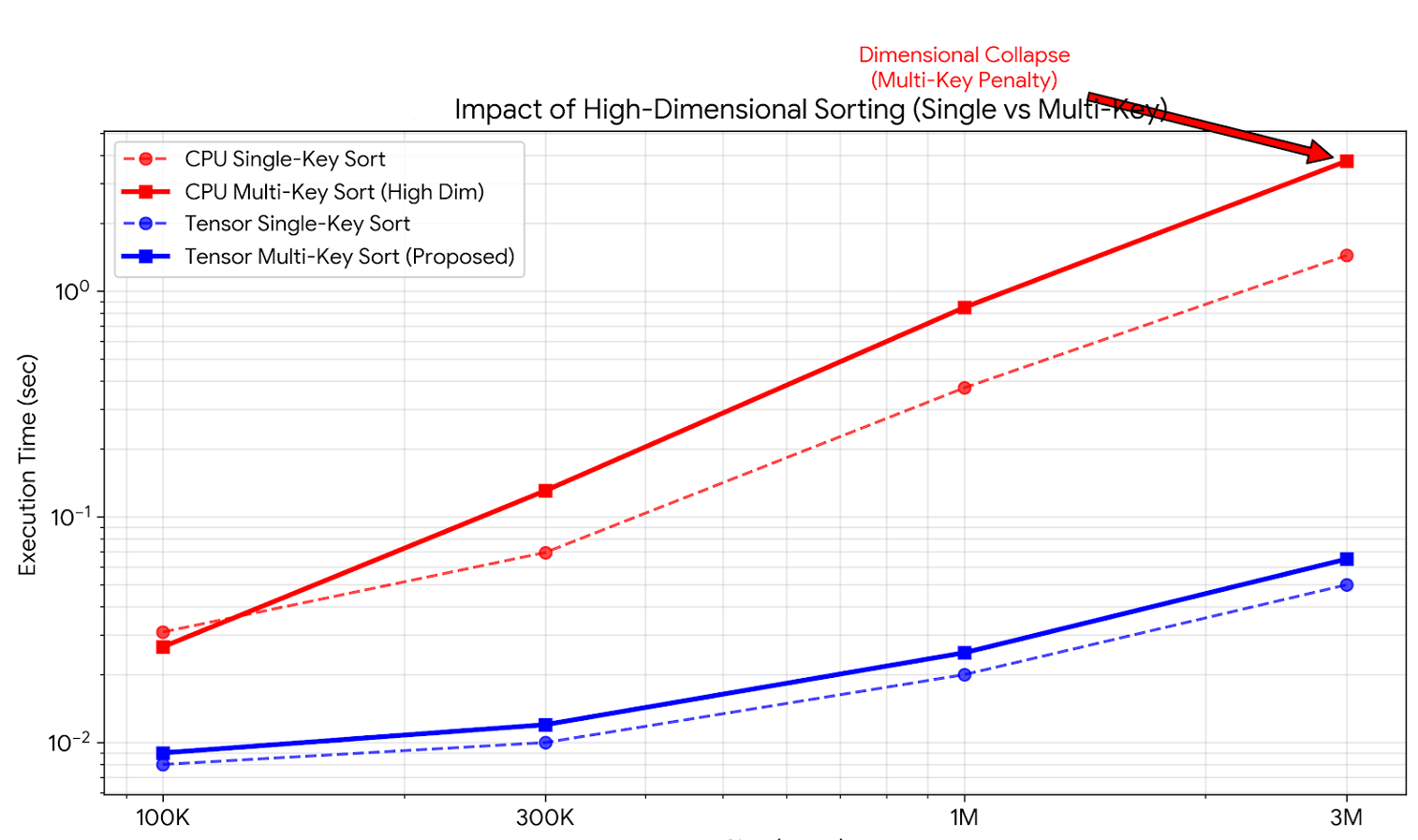}
    \caption{Impact of High-Dimensional Sorting (Single vs Multi-Key)}
\end{figure}

Under the condition of input N=1,000,000 and work\_mem=1MB, the relational execution path generated about 200.41MB of temporary disk I/O, and P99 latency exceeded 2 seconds. In contrast, the tensor-based execution path maintained a stable P99 latency of about 0.56 seconds without physical spilling under the same conditions.

\subsection{Execution-time Path Selection Analysis}
To evaluate the effect of execution-time path selection, results from forcibly applying either the linear execution path or the tensor-based execution path were compared with results from applying execution-time selection on the same query and dataset. As a result, when execution-time selection was applied, it was confirmed that the worst execution time could be effectively avoided by selecting tensor-based execution after the intersection point. Figure 6 shows a comparison between the P99 latency of CPU hash joins according to work\_mem settings and tensor-based execution.

\begin{figure}[h]
    \centering
    \includegraphics[width=0.8\linewidth]{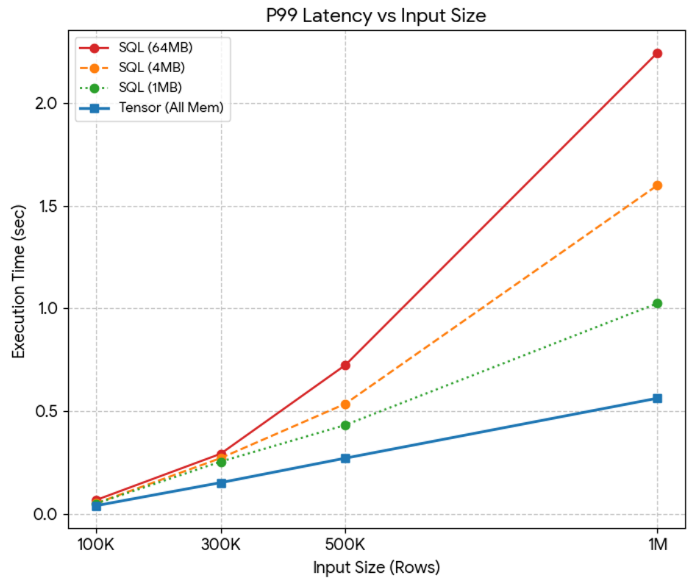}
    \caption{P99 Latency vs Input Size}
\end{figure}

Figure 7 shows the temporary disk usage under the same conditions. CPU-based execution paths show a sharp increase in spilling when memory is insufficient, which is directly connected to increased execution time. These observations motivate a structural interpretation of this behavior as a memory-regime shift driven by spill amplification.

\begin{figure}[h]
    \centering
    \includegraphics[width=0.8\linewidth]{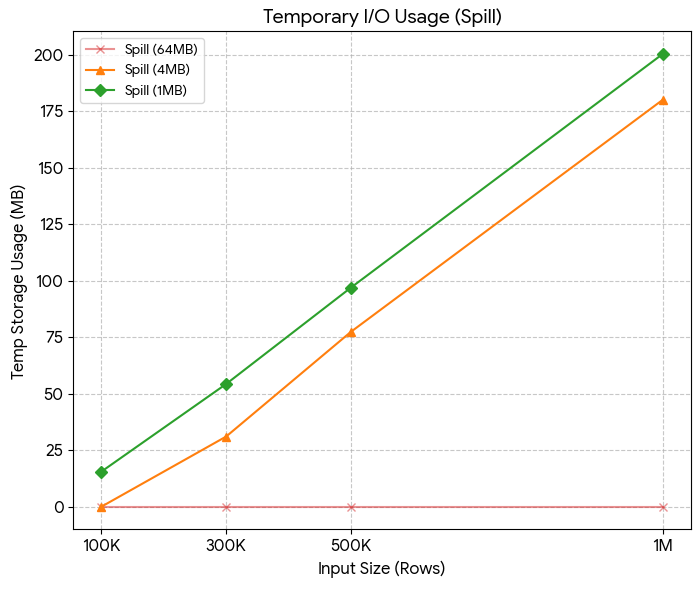}
    \caption{P99 Temporary I/O Usage (Spill)}
\end{figure}

\section{Structural Cost Interpretation (Regime Shift \& Predictability)}
\label{sec:structural}

The experimental results indicate that the observed tail-latency inflation is not merely an implementation artifact, but a structural consequence of premature dimensional collapse in tuple-oriented execution.

Let $N$ be the input size and $M$ be the effective memory budget (e.g., \texttt{work\_mem}). In relational hash join execution, intermediate structures (hash tables and partitions) must be materialized in a linearized form. When the working set exceeds $M$, the operator transitions into a spill regime, repeatedly partitioning and writing temporary runs to disk. This introduces an additional recovery cost beyond the ideal compute-bound scaling.

We model this transition as a regime shift:
\[
T_{\text{rel}}(N) = O(N) + \alpha(N, M),
\]
where $\alpha(N, M)$ denotes the spill amplification cost induced by repeated partitioning and temporary I/O. This formulation highlights that the instability is not algorithmic inefficiency but a representation-induced structural overhead emerging under memory constraints. Importantly, $\alpha(N, M)$ grows superlinearly as memory pressure increases, since both the number of partitions and the volume of re-materialized data expand as a function of the memory deficit.

This regime shift directly explains predictability loss. Under memory pressure, the relational execution path becomes sensitive to transient I/O scheduling and partitioning variance, causing the latency distribution to widen (P50--P99 dispersion). Empirically, at $N=1{,}000{,}000$ and \texttt{work\_mem=1MB}, the relational path generated approximately 200.41MB of temporary spill and exhibited P99 latency exceeding 2 seconds, consistent with spill-bound behavior.

In contrast, the tensor-based execution path preserves more aggregated structural representations and delays or bypasses premature dimensional collapse. As a result, it avoids entering the spill amplification regime and maintains a more deterministic scaling profile:
\[
T_{\text{tensor}}(N) \approx O(N).
\]
Therefore, the key advantage of tensor-based execution is not acceleration alone, but structural resilience: while relational execution loses predictability under memory pressure, tensor-based execution maintains more deterministic behavior by preventing spill amplification.

Unlike prior GPU-accelerated database systems that primarily aim to improve raw throughput \cite{bress2014, he2008}, the tensorized path proposed here targets structural predictability under constrained memory regimes, focusing on tail-latency stability rather than peak performance alone. This perspective complements prior work on risk-aware cardinality estimation, which addresses plan selection uncertainty at optimization time, whereas this work addresses structural execution instability at runtime.

\section{Conclusion}
This study reinterpreted performance instability in cost-based relational execution as a structural consequence of premature dimensional collapse, and demonstrated that execution-time representation choice can improve predictability under memory pressure. Relational execution paths flatten data at an early stage of operation and rely on hash-based structures to perform joins. Experimental results quantitatively support these structural characteristics. Under the condition of input N=1,000,000 and work\_mem=1MB, the relational execution path generated approximately 200.41MB of temporary disk I/O, and the P99 tail latency exceeded 2 seconds. In contrast, the tensor-based execution path maintained a stable P99 latency of about 0.56 seconds without physical spilling. While relational execution loses predictability under memory pressure, tensor-based execution maintains more deterministic expansion characteristics. Designing execution paths that delay or bypass premature dimensional collapse can be a key strategy for ensuring the stability of tail latency and reducing structural recovery costs. Future work will investigate adaptive integration of tensorized execution with cost-based plan selection under varying data distributions and join selectivities.

\section*{Acknowledgments}
The author used AI-assisted tools (e.g., ChatGPT) for English grammar correction and language polishing. All technical content, experiments, and conclusions were developed and verified by the author.

\end{document}